\documentclass[aps,twocolumn,showpacs,superscriptaddress,amsmath,amssymb,amsfonts,floatfix,longbibliography]{revtex4-1}

\usepackage[version=3]{mhchem}
\usepackage[T1]{fontenc} 
\usepackage{graphicx}
\usepackage{dcolumn}
\usepackage{bm}
\usepackage{amssymb}
\usepackage[dvipsnames]{xcolor}
\usepackage{microtype}
\usepackage{xfrac}
\usepackage{gensymb}
\usepackage[most]{tcolorbox}
\usepackage{xcolor}
\usepackage{multirow}
\usepackage{enumitem}
\usepackage{natbib,hyperref}
\usepackage{graphicx}

\usepackage{color}

\hypersetup{
	colorlinks=true,
	linkcolor=blue, 
        citecolor=Aquamarine,
}

\begin{document}

\title{Understanding the role of entropy in high entropy oxides} 

\author{Solveig S. Aamlid}
\affiliation{Stewart Blusson Quantum Matter Institute, University of British Columbia, Vancouver, BC V6T 1Z4, Canada}

\author{Mohamed Oudah}
\affiliation{Stewart Blusson Quantum Matter Institute, University of British Columbia, Vancouver, BC V6T 1Z4, Canada}

\author{J\"org Rottler}
\affiliation{Stewart Blusson Quantum Matter Institute, University of British Columbia, Vancouver, BC V6T 1Z4, Canada}
\affiliation{Department of Physics \& Astronomy, University of British Columbia, Vancouver, BC V6T 1Z1, Canada}

\author{Alannah M. Hallas}
\email[Email: ]{alannah.hallas@ubc.ca}

\affiliation{Stewart Blusson Quantum Matter Institute, University of British Columbia, Vancouver, BC V6T 1Z4, Canada}

\affiliation{Department of Physics \& Astronomy, University of British Columbia, Vancouver, BC V6T 1Z1, Canada}

\begin{abstract}
The field of high entropy oxides (HEOs) flips traditional materials science paradigms on their head by seeking to understand what properties arise in the presence of profound configurational disorder. This disorder, which originates from multiple elements sharing a single lattice site, can take on a kaleidoscopic character due to the vast numbers of possible elemental combinations. High configurational disorder appears to imbue some HEOs with functional properties that far surpass their non-disordered analogs. While experimental discoveries abound, efforts to characterize the true magnitude of the configurational entropy and understand its role in stabilizing new phases and generating superior functional properties have lagged behind. Understanding the role of configurational disorder in existing HEOs is the crucial link to unlocking the rational design of new HEOs with targeted properties. In this Perspective, we attempt to establish a framework for articulating and beginning to address these questions in pursuit of a deeper understanding of the true role of entropy in HEOs. 

\end{abstract}
\maketitle

\section{Introduction}
Disorder and defects, in low concentrations, can both generate and suppress functional materials properties. For example, silicon in technological applications is famous for its exquisite purity, but without a finite level of substitutional disorder, it would not exhibit the requisite electronic properties needed for its widespread use as the backbone of all modern electronics~\cite{spear1975substitutional,madou2018fundamentals}. In the opposite limit, quantum materials that are currently being examined for next generation technological applications can have their magnetic or electronic states disrupted by astonishingly low levels of vacancies~\cite{phelan2016chemistry} or site mixing~\cite{arpino2017impact}. In these examples, disorder is a weak perturbation on an otherwise highly crystallographically ordered and compositionally clean material. The burgeoning field of high entropy materials flips this paradigm on its head by seeking to understand what material properties can emerge in the limit of extreme disorder - where the elements are mixed in roughly equiatomic ratios and disorder is no longer a weak perturbation but is instead one of the leading energy scales at play.

High entropy materials first rose to prominence in the early 2000s following two seminal papers by Yeh and Cantor on metallic alloys with five or six principal components \cite{Cantor2004, Yeh2004}. These alloys immediately struck a chord with the materials science community as they were shown to have enhanced mechanical properties, and have shown promise for a wide range of potential applications such as refractory~\cite{Senkov2010}, catalytic~\cite{yao2018carbothermal}, and cryogenic~\cite{gludovatz2014fracture} technologies, among many others. As the field has grown, a number of terms have been introduced to describe these types of alloys including multi-component alloys, compositionally complex alloys, and multi-principal-element alloys. Over time, however, the term \textit{high entropy alloy} (HEA) has remained most prominent. That language has now been adopted to describe many other classes of materials including high entropy oxides, carbides, and borides, among others.

\begin{figure*}[!ht]
  \centering
  \includegraphics[width=17cm]{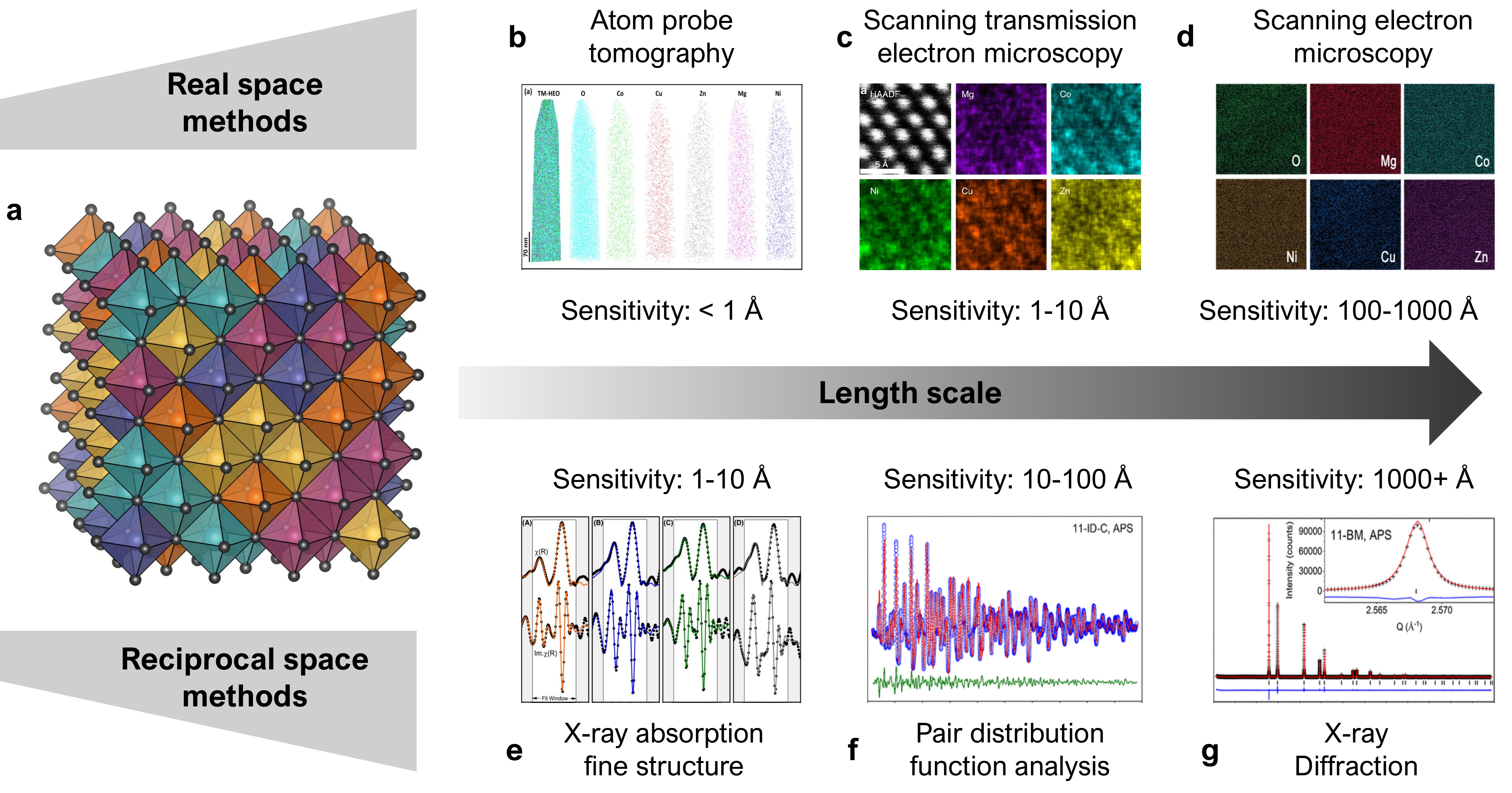}
\caption{(a) High entropy oxides (HEOs) are characterized by a random distribution of metal cations across an ordered crystalline lattice, as represented here for five different metals on the rock salt lattice, illustrated using VESTA~\cite{momma2008vesta}. Experimental verification of these characteristics requires a range of (b-d) real space and (e-g) reciprocal space methods, with sensitivity to different length scales. All of the representative data shown in this figure were collected on the prototypical rock salt HEO (Mg,Co,Ni,Cu,Zn)O. (b) Reprinted from \cite{chellali2019homogeneity} with permission from Elsevier; (c) Reprinted from~\cite{Rost2015} under Creative Commons CC-BY license; (d) Reproduced from \cite{hong2019microstructural} with permission from Wiley; (e) Reproduced \cite{rost2017local} with permission from Wiley; (f) and (g) reproduced from \textit{Chem. Mater.} 2019, 31, 10, 3705–3711. \copyright~2019 American Chemical Society~\cite{zhang2019long}.}
    \label{fig:HEO_Lengthscales}
\end{figure*}

There is no universally agreed upon singular definition for what constitutes a high entropy material. However, there are a handful of widely agreed upon characteristics~\cite{miracle2017critical,sarkar2020high}. The first is that high entropy materials are crystalline, meaning the atoms occupy a crystal lattice with well-defined space and point group symmetries. Therefore, the entropy is not primarily related to positional disorder, as would be the case for an amorphous material. The second trait is that they should exist as a single phase material rather than a mixture of phases. Finally, a high entropy material should have significant configurational disorder due to multiple elements inhabiting the same crystallographic site. These elements should be randomly distributed on all length scales, all the way down to the atomic scale. Demonstrating that a material meets these three criteria necessitates that high entropy materials be studied over many orders of magnitude in length-scale, as schematically shown in Figure \ref{fig:HEO_Lengthscales}.

Canonically, materials in which multiple elements share a single crystallographic site are known as solid solutions. Therefore, distinguishing a high entropy material from a solid solution requires some further criterion, and this is where consensus is lacking. Quantitatively, one might hope to distinguish a high entropy material by the number of elements involved in its solid solution and therefore the magnitude of its configurational entropy, while qualitatively, one might require that a high entropy material should force in a crystal structure distinct from a subset of its precursors. Experimentally, one might demonstrate that the material forms reversibly with an endothermic reaction enthalpy, while theoretically one might show the material to have a positive enthalpy of formation suggestive of the decisive role of entropy.  Each of the criteria, discussed in Section II, have merit, but none individually encompasses all materials that have been included in the umbrella of \textit{high entropy}.

While many of the topics discussed in this Perspective will be broadly applicable to all classes of high entropy materials, our examples will primarily come from the oxides. There are a number of advantages to studying high entropy oxides (HEOs): Oxides, in comparison to alloys, exist in a wider variety of crystal structures and local site symmetries, they are generally stable, their interactions are mediated by the oxygen sublattice, and they are technologically important. HEOs have seen an explosion of interest following a seminal 2015 work by Rost \textit{et al.}, reporting the discovery of a five-component rock salt material (Mg,Co,Ni,Cu,Zn)O whose single phase crystal structure was claimed to be stabilized by entropy \cite{Rost2015}. As was the case with HEAs, the HEOs immediately received significant attention in the materials science community, and very quickly they were shown to have promising characteristics for a wide range of applications, including enhanced stability through charge cycling for battery applications~\cite{sarkar2018high}. 

The pivotal question facing this field is one that goes far beyond reaching a consensus on the precise definition of high entropy. Simply put, the question is \textit{what is the role of entropy in high entropy oxides?} To be more specific: How and when does entropy contribute to the stabilization of new phases, what are the experimental or theoretical benchmarks that we can use to distinguish an HEO from a non-HEO, to what extent do other factors enhance or suppress the ideal configurational entropy, and ultimately, what role does this entropy play in the materials properties that emerge at temperatures far below the synthesis temperature? The goal of this Perspective will be to establish a framework for articulating and beginning to address these questions. We hope that this Perspective article will complement a number of excellent existing reviews that have surveyed known HEO materials~\cite{musico2020emergent,oses2020high}, discussed their design principles and synthesis~\cite{zhang2019review}, addressed their thermodynamics~\cite{musico2020emergent,sarkar2020high,mccormack2021thermodynamics}, and examined their functional properties~\cite{zhang2019review,oses2020high,sarkar2020high,musico2020emergent,toher2022high}.

\section{What defines a high entropy material?}

One problem facing the field of high entropy materials is linguistic in nature -- what defines a material as \emph{high entropy} in the first place? Terminology in this field has continually evolved and differing criteria for what constitutes a high entropy material have been put forward over the years, but these have at times been applied inconsistently~\cite{brahlek2022name}. In the Introduction, we outlined crystallinity, configurational disorder, and phase purity as necessary but not sufficient conditions to identify a high entropy material. It is worth noting that the alloys to which the high entropy label was first applied did not completely fulfill any of these three criteria - they were phase separated, elementally inhomogeneous, and contained amorphous regions~\cite{Yeh2004,Cantor2004}. However, later iterations of the same compositions fulfilling all these criteria were subsequently prepared upon identification of the appropriate synthesis conditions~\cite{cantor2021multicomponent}. In oxide materials more specifically, the three requisite traits outlined above are fulfilled by many conventional solid solutions and therefore some additional criteria is required to differentiate HEOs. In this section, we will discuss some of the common ways high entropy materials have been distinguished from conventional solid solutions in the literature and attempt to identify the advantages and disadvantages associated with each.

\subsection{Configurational entropy larger than $1.5R$}

\begin{figure}
    \centering
    \includegraphics[width=8.5cm]{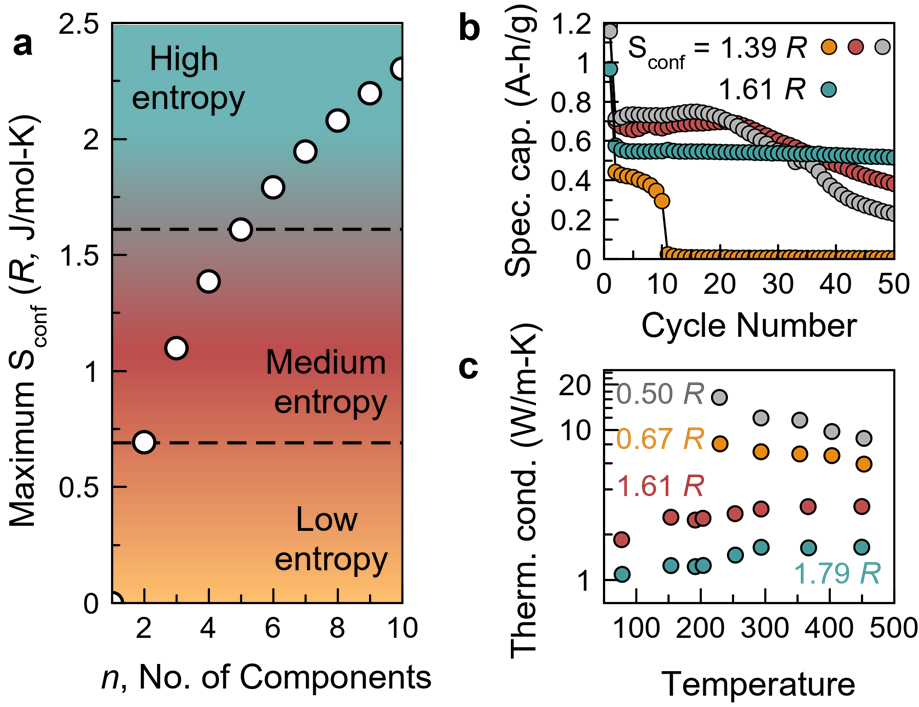}
    \caption{(a) Ideal configurational entropy as a function of number of constituents, $n$, in equiatomic proportions showing the boundary between low and medium entropy at $S_{conf}=0.69R$ and the boundary between medium and high at $S_{conf}=1.61R$. This definition of high entropy can only be achieved with a minimum of five constituent elements sharing a crystallographic site. Two examples of enhanced functionality with constituent number are shown in (b) for the cycling stability for Li-ion battery applications, redrawn from Ref.~\cite{sarkar2018high} and (c) the suppression of thermal conductivity with possible applications in thermal barrier coatings, redrawn from Ref~\cite{braun2018charge}.}
    \label{Entropy_calc}
\end{figure}

Early attempts to define high entropy materials considered the number of constituents and their relative stoichiometries. Starting from Boltzmann's fundamental entropy formula $S=k_B \ln{\Omega}$, one can enumerate the number of microstates $\Omega$ associated with the configurational entropy of multiple metal ions occupying the same crystallographic site, $S_{conf}$. Considering first the number of possible distributions of $n_i$ particles of type $i$ over a total number of $N=\sum_i n_i$ states yields the combinatorial factor $\Omega=N!/\prod_i n_i!,$ which results in the \emph{ideal} entropy
\begin{equation}
    S_{conf}=R\ln\left[\frac{N!}{\prod_i n_i!}\right]\approx -R \sum_i x_i \ln{x_i}
\label{sid-eq}
\end{equation}
where $x_i=n_i/N$ and we have used Stirling's approximation. For any number of constituents, this configurational entropy is maximized when the elements are found in equiatomic proportions, as shown for numbers of constituents between 1 and 10 in Fig.~\ref{Entropy_calc}(a). In the case of oxides, this definition can be straightforwardly adapted to include a second sum over sublattices 
\begin{equation}
    S_{conf}= -R \sum_s m_s \sum_i x_{i,s} \ln{x_{i,s}},
\label{sublattice-eq}
\end{equation}
where $m_s$ is the multiplicity of sublattice $s$ and $x_{i,s}$ is the fraction of element $i$ on sublattice $s$. This modification accounts for potential contributions to the total configurational entropy arising from multiple cation sublattices, oxygen vacancies, or other forms of disorder on the anion sublattice. 

Empirical observations from the alloy field led to the conclusion that there was a special effect upon crossing the threshold from four to five elements, as this is when a single phase material could generally be obtained~\cite{Yeh2004,Cantor2004}. Therefore, an early definition for a high entropy material was one in which the \emph{ideal} configurational entropy meets or exceeds the value for five elements in equiatomic proportions, which is $S_{conf}^{high} \gtrsim 1.61R$~\cite{jien2006recent}. The corresponding definition for low entropy was taken as the maximum value for two components, which is $S^{low}_{conf} \lesssim 0.69 R$ and these are both indicated in Fig.~\ref{Entropy_calc}(a). Materials falling intermediate to these two extremes could then be considered as medium entropy. Other works have rounded down the requirement for high entropy to any material with $S^{high}_{conf} \geq 1.5R$ and rounded up the low entropy cutoff to $S^{low}_{conf} \leq 1.0R$~\cite{murty2019high}. This criterion for high entropy has been directly carried over to the oxide field.

While this definition is satisfying in its simplicity, it fails to account for the true complexity of real materials. Equation (\ref{sid-eq}) presupposes a truly random distribution of cations, yielding the maximum, ideal configurational disorder. The experimental reality is undoubtedly less ideal. Inhomogeneity, whether due to kinetic barriers from an initially non-uniform distribution of ions or due to energetically preferred clustering or short-range ordering would immediately reduce the entropy from its ideal value. While a few studies have demonstrated a very high degree of randomness at the atomic scale (see for example Fig.~\ref{fig:HEO_Lengthscales}(b,c)), these cannot rule out hyper-local short range ordering. In particular, one topic that has not yet received sufficient investigation is the extent to which pairwise preferences may act to create extremely local short range ordering that reduces the configurational entropy below the simple $x\ln x$ description - a topic we will return to in Section \ref{fac-reducing-section}.

A further point of interest is the appropriateness of the $S^{high}_{conf} \geq 1.5R$ threshold for oxides, which was initially established for alloys with a single sublattice. In seeking to understand whether this boundary is in need of deeper scrutiny, we can consider whether the entropy required to stabilize a structure in the alloys differs significantly from what is required to stabilize an oxide structure. Our survey of calorimetric enthalpies for intermetallics~\cite{kim2017experimental} (with the reactants being metals and the products being primarily binary intermetallics) and for oxides~\cite{hautier2012accuracy} (with the reactants being binary oxides and the products being ternary oxides in order to exclude the typically very large enthalpy of oxidation) reveals that the average enthalpy change in these databases is 492 meV/cation for the oxides and 468 meV/atom for the intermetallics. While enthalpy is not the only consideration, the similarity of these values supports the notion that the $1.5R$ boundary has not been misappropriated in the oxides. 

There are some compelling examples where the $S^{high}_{conf} \geq 1.5R$ indeed appears to act as a threshold for the enhancement of functional properties. For instance, an investigation of the electrochemical properties of the rock salt HEO for Li-ion battery applications showed a pronounced enhancement of the capacity retention upon cycling for the five-component material as compared to any of the four-component analogs, as shown in Fig.~\ref{Entropy_calc}(b)~\cite{sarkar2018high}. Along similar lines, another study of the five-component rock salt material showed that this compound exhibits highly favorable thermal and mechanical properties for applications and that the thermal conductivity can be suppressed by an additional factor of two by introducing a sixth constituent element, as shown in Fig.~\ref{Entropy_calc}(c)~\cite{braun2018charge}. The exact mechanism behind such ``cocktail'' effects remains an open question but it is clear that, in some cases, greater stability is imbued in the more configurationally disordered phases, particularly those exceeding $S^{high}_{conf} \geq 1.5R$.

\subsection{Entropy stabilization}

An alternative means of distinguishing a high entropy materials from a conventional solid solution is the requirement that the material itself is stabilized by entropy. Entropy stabilization occurs when entropy dominates the thermodynamic landscape at the synthesis temperature and plays the decisive role in either stabilizing a single phase material, choosing the resulting crystal structure, or both. Given the Gibbs free energy equation $G = H - TS$, one can immediately appreciate that entropic contributions to the free energy grow commensurately with temperature. In contrast, the enthalpy of formation is almost temperature independent and therefore defines the true ground state in the zero temperature limit. This reveals a key feature of entropy stabilized materials: they are typically only thermodynamically stable at elevated temperatures. Therefore, in the synthesis process one must often rapidly quench the material in order to kinetically trap the metastable phase. 

The term \textit{entropy stabilization} was first used in 1968 by Navrotsky and Kleppa when they observed positive enthalpies of formation for a subset of spinels in drop solution calorimetry and theorized that the reaction occurred because of the entropy contribution from site-mixing between the two cation sites \cite{Navrotsky1968}. In the modern sense, the concept of entropy stabilization originated from the observation that increasing the number of elements in an alloy resulted in a higher chance of forming a single phase \cite{Yeh2004}. An advantage of thinking in terms of entropy stabilization is that there are numerous direct experimental and theoretical observables that can provide strong evidence for or against this scenario, as summarized in Figure \ref{Entropy stabilization}:

\begin{figure}
    \centering
    \includegraphics[width=8.5cm]{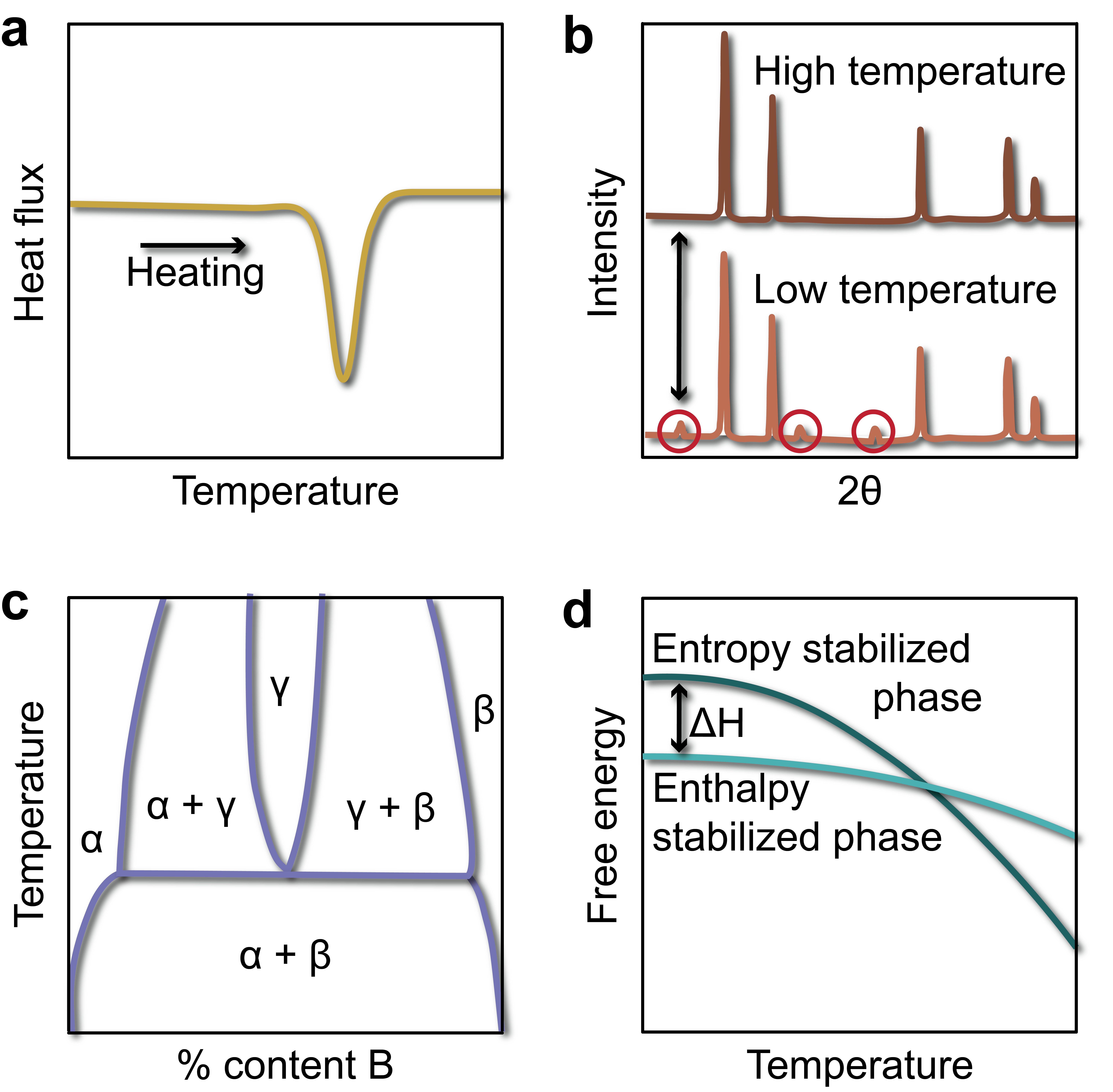}
    \caption{Entropy stabilization can be justified using multiple approaches such as (a) calorimetric verification of endothermic heat of formation, (b) demonstrating reversibility of the phase transformation, exemplified by the formation of impurity peaks in x-ray diffraction after heat treatments below the phase transition, (c) the formation of a phase in the center of the phase diagram dissimilar to any phases at the edges of the phase diagram, ideally only at high temperatures, and  (d) calculating the $\Delta H$ of formation of two competing entropy stabilized and enthalpy stabilized phases using DFT and ensuring the configurational entropy will dominate the free energy at some temperature below melting.}
    \label{Entropy stabilization}
\end{figure}

\begin{itemize}[leftmargin=*]

\item \textbf{Endothermic reaction enthalpy:} Following the thermodynamic arguments laid out above, it follows that in order for a material to be considered entropy stabilized, the reaction enthalpy of the competing binary oxide phases should be positive. In such a scenario, the chemical reaction can proceed only when the unfavorable change in enthalpy is overcome through the absorption of sufficient heat from the environment, resulting in the formation of the entropically selected state. Therefore, one of the most direct signatures of entropy stabilization is that the chemical reaction from the constituent oxides to the single phase material should proceed endothermically. Direct experimental measures of reaction enthalpies can be accomplished by various calorimetric techniques, including differential scanning calorimetry, whose sign will directly indicate whether the phase transition is endothermic, as shown schematically in Figure~\ref{Entropy stabilization}(a). It is important to emphasize here that configurational entropy is not the only contribution to the entropic term; in particular, the vibrational entropy, which will be introduced in Section~\ref{Contributions-to-S} below, is also a significant contribution. Configurational entropy can, however, play the decisive role in tipping the balance~\cite{mccormack2021thermodynamics}.

\item \textbf{Observation of a reversible phase transition:}
Reversibility -- the phenomena whereby a chemical reaction under some set conditions proceeds in both a forward and reverse direction simultaneously -- has been put forward as one of the gold standard experimental signatures of entropy stabilization. The reason for this can again be understood by considering the Gibbs free energy relation, $G = H - TS$, where we see that entropy stabilization can only occur when the temperature is sufficiently large to drive the formation of a single phase material. If the material is not quenched to thermodynamically trap it then there should be some temperature threshold below which the entropy term becomes too small to dominate the free energy landscape, at which point the material will decompose into multiple enthalpy preferred phases. The point of reversibility therefore occurs when the temperature is such that the enthalpy and entropy terms are exactly balanced.

In order to experimentally demonstrate reversibility in favor of a scenario of entropy stabilization, one should therefore (i) demonstrate the formation of a single phase material through quenching from the synthesis temperature, (ii) demonstrate that slowly cooling or annealing the material at some lower temperature than the initial synthesis temperature, results in a partial or complete decomposition of the single phase, and finally (iii) demonstrate that re-annealing the material at the synthesis temperature followed once again by rapid quenching re-stabilizes the single phase material. This process is demonstrated in Fig.~\ref{Entropy stabilization}(b), using x-ray diffraction to detect the formation of a single phase at high temperatures and the presence of impurity peaks at low temperature. There are, however, a handful of issues with using reversibility as the sole determinant of entropy stabilization. First, it is not guaranteed that a truly entropy stabilized material will demonstrate reversibility on any reasonable time scale if the kinetic barrier to transform from its metastable state is too large, (see also Section \ref{sec-kinetics}). Second, reversibility is not an unambiguous indicator of entropy stabilization and can instead be reflective of passing through a solubility threshold, as will be discussed further below. Experiments to demonstrate reversibility are certainly worthwhile and can be informative of the energy landscape in which a material exists; however it is in-and-of-itself neither a necessary nor a sufficient condition on which to establish entropy stabilization. Furthermore, reversibility is undoubtedly an undesirable trait when considering applications if it occurs on any timescale at the operation temperature, as the material will inevitably degrade, negating any benefits in its functional properties.

\item \textbf{The formation of a crystal structure distinct from any end-member:} Empirically, one useful indicator that a given solid solution may be entropy stabilized is if it is distinct from any or most of the structures of the precursors. For instance, in the case of the aforementioned rock salt HEO, with chemical formula (Mg,Co,Ni,Cu,Zn)O, each of MgO, CoO, and NiO is itself a rock salt material while CuO and ZnO form in the lower symmetry tenorite and wurtzite structures, respectively, meaning that 40\% of the cations occupying the rock salt HEO phase are in higher energy structures relative to their ground states. A more extreme example comes from the solid solution of \ce{TiO2} and \ce{ZrO2}, which independently are examples of tetragonal rutile and mononclinic baddeleyite structures. When mixed in close to equiatomic ratios and reacted at high temperatures, \ce{TiZrO4} forms a solid solution with the orthorhombic $\alpha$-PbO$_2$ structure \cite{Troitzsch2005}. A generalized depiction of what this may look like in a phase diagram is shown in Fig.~\ref{Entropy stabilization}(c). This is a rare example where 100\% of the cations are in a non-ground-state crystal structure, which is clear-cut evidence for the strong role of entropy. To our knowledge, there are no reported examples of such systems with five or more constituents but (Ti,Zr,Hf,Sn)O$_2$, also with the $\alpha$-\ce{PbO2} structure, is a four-component example~\cite{AamlidinProgress}. Along similar lines, if removing a component and thereby reducing the entropy leads to the formation of a non-single phase material, that too is generally indicative that entropy may be playing a leading role. 

\item \textbf{Theoretical determination of a positive enthalpy of formation:} Density functional theory (DFT) provides a theoretical tool to calculate the enthalpy of formation of any chemical composition in any crystal structure relative to a ground state from first principles. In the case of HEOs, the enthalpy of reaction between multiple ordered phases and one high entropy disordered phase can be determined, and the viability of a reaction can be assessed by comparing the calculated enthalpy with the ideal entropy of mixing to find a hypothetical transition temperature. This is illustrated in Figure \ref{Entropy stabilization}(d). The accuracy of DFT calculations for high entropy materials may not be precise enough to predict what crystal structure will form. Nonetheless, the ease of calculation compared to synthesis and calorimetry measurements makes it a useful tool in determining what combinations of elements might be viable, and can give evidence towards a positive formation enthalpy when experimental tools are not available or practical.

\end{itemize}

The rock salt HEO (Mg,Co,Ni,Cu,Zn)O is widely considered the gold standard for a clear cut example of configurational entropy stabilization among five-component oxides. It exhibits each of the above mentioned experimental and theoretical hallmarks of an entropy stabilized phase. However, a study by Fracchia \emph{et al.} has cast doubt on the leading role of configurational entropy in stabilizing the resulting rock salt phase~\cite{fracchia2022configurational}. Their study demonstrates that the phase behavior of the five-component rock salt is not markedly different than that of a three-component rock salt with the same ratio of phases that have a pre-existing rock salt ground state (MgO, CoO, and NiO) vs. those without (ZnO and CuO). This study implicates the prominent role of conventional solubility limits, where dissolution processes are, indeed, generally endothermic and therefore entropically-driven.

The picture arising following this study makes it clear that disentangling configurational entropy stabilization and conventional solid solution solubility limits is a challenging endeavor. Indeed, the experimental observables of a reversible transition between a mono-phase and bi-phase region of a phase diagram and the resulting metastable quenched phase are nearly identical to those of a  true configurational entropy stabilized material. Therefore, a new gold standard for a definitive example of stabilization by configurational entropy might be the experimental demonstration that the decomposition of the single phase material varies in some systematic way with the configurational entropy as controlled by the number of constituents.

\subsection{Final remarks on high entropy nomenclature}

\begin{figure}
    \centering
    \includegraphics[width=8.5cm]{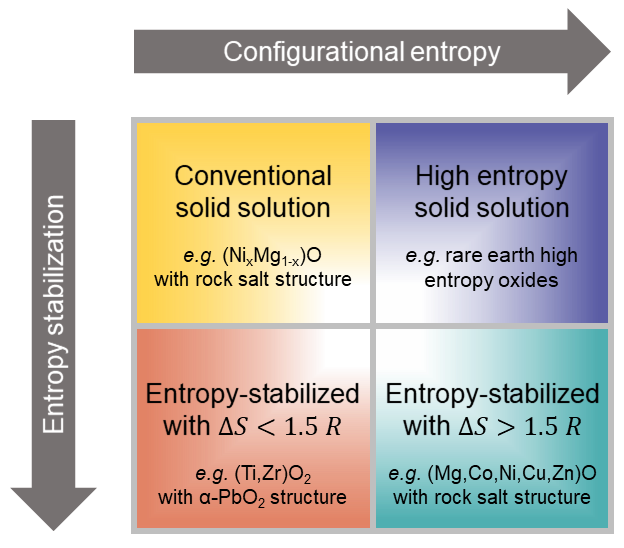}
    \caption{HEOs can be categorized according to two related but distinct characteristics: the magnitude of their configurational entropy and the role of entropy in stabilizing their observed crystal structure. This is schematically represented with phases with increasing configurational entropy shown from left to right and the degree of entropy stabilization increases going from top to bottom. Known materials can be sorted into each of the four resulting quadrants. Conventional solid solutions have low configurational entropy and are not entropy stabilized while HEOs exhibit either configurational entropy exceeding $1.5R$ and/or entropy stabilization. }
    \label{MaterialsGrid}
\end{figure}

While we cannot endorse any one of the above discussed features as the definitive criteria for an HEO, we nonetheless wish to conclude this section by stressing the need for precision in describing the entropic characteristics of a given material. This is summarized in Figure~\ref{MaterialsGrid}, where we schematically organize materials according to two uncorrelated axes, increasing configurational entropy and increasing entropy stabilization. While what we have termed conventional solid solutions are low in both metrics, HEOs might satisfy the high threshold in either one or both of these characteristics. For instance, it is important to emphasize that a material can have a high configurational entropy, exceeding the empirical threshold of $S_{conf} \geq 1.5R$ without being entropy stabilized. Solid solutions in which a single crystalline site is occupied by multiple rare earth ions will almost always fall into this category. Due to their highly localized valence electrons, rare earths are chemically almost indistinguishable from one another, which is why they are so difficult to separate in natural ores. The driving force of configurational entropy therefore need not be invoked to understand why rare earths recombine to form oxide solid solutions. Furthermore, it is often the case that each rare earth independently will form in the same structure as the resulting high entropy phase and therefore the structure itself cannot be viewed as entropically selected. Rare earth solid solutions therefore occupy the upper right quadrant of our schematic shown in Figure~\ref{MaterialsGrid}.
 
Likewise, a material can also certainly be entropy stabilized without exceeding the $1.5R$ threshold, as is the case for the two-component spinels studied by Navrotsky in the late 1960s~\cite{Navrotsky1968}. This scenario is represented by the lower left quadrant in Figure~\ref{MaterialsGrid}. The above mentioned (Ti,Zr)O$_2$ and (Ti,Zr,Hf,Sn)O$_2$ with the $\alpha$-\ce{PbO2} structure are also unambiguously entropy stabilized, with their endothermic formation~\cite{Hom2001} and crystal structure that is distinct from any of the constituent oxides. These materials with two and four constituents are respectively, examples of low- and medium-entropy oxides. However, given the prominent role of entropy in their formation, we argue that it is nonsensical to exclude them from the umbrella of HEOs. Supporting this view, in a recent survey of the metal alloy systems, it was found that the enthalpy contribution to the stabilization of an ordered N component phase relative to its N-1 competing phases is small once the ordered compound contains three elements or more \cite{Toher2019}. This means that entropy dominates over enthalpy already at three constituent elements, and the empirical threshold at five elements might not be appropriate for all systems. Thus, we can conclude that the linguistic confusion in the HEO field is a natural consequence of the observation that no material appears to completely fulfill all possible criteria and the only remedy is to clearly articulate the nature of the HEO for any given material.

\section{Symmetry and symmetry breaking in high entropy oxides}

Crystallinity is one of the core requirements for a high entropy material, meaning it should exist in a crystal structure whose symmetry properties are encoded in one of the 230 possible space groups for 3-dimensional solids. Interestingly, an informal survey of known high entropy materials reveals a decided tendency towards high symmetry structures, primarily those with cubic space groups. For instance, high entropy alloys with body-centered cubic (BCC) and face-centered cubic (FCC) structures are overwhelmingly represented~\cite{george2019high}. Similarly, for the HEOs the cubic rock salt, fluorite, perovskite, and spinel structure types are the most prominent~\cite{musico2020emergent,mccormack2021thermodynamics}. Part of this observation can almost certainly be accounted for by the underlying frequency distribution for the natural occurrence of space groups in solids, which is itself known to be highly uneven. The general tendency does skew towards higher symmetry -- for example, all of the 10 most common space groups accounting for more than two thirds of inorganic solids, belong to centrosymmetric and holohedral space groups~\cite{urusov2009frequency}. Yet only two of the top ten most common space groups are cubic. In contrast, nearly 80\% of the compounds surveyed in a 2020 review of HEOs belonged to cubic space groups~\cite{musico2020emergent}, suggestive of their strong overrepresentation. In fact, almost all examples where the symmetry is reduced from cubic contained a second \textit{ordered} cation sublattice such as the orthorhombic perovskites $R$(Cr,Mn,Fe,Co,Ni)O$_3$, where $R$ is a rare earth element~\cite{sarkar2018rare,sharma2020magnetic}. Still, for the majority of known HEOs, it appears that there is a strong preference towards higher symmetry, and especially cubic, structures.

The observed frequency of high symmetry crystal structures for HEOs can, in part, be understood on intuitive grounds. Similarly to how most pure metals and high entropy alloys are crystallizing in close-packed structures, the large oxygen anion in an oxide will tend to form a close-packed lattice with the smaller cations filling in voids. Pauling’s fifth rule ‘The number of essentially different kinds of constituents in a crystal tends to be small’ would indicate that high symmetry structures with one kind of regular polyhedron is generally preferred~\cite{Pauling1929}, and indeed the majority of oxides do exhibit highly symmetric local oxygen environments. The close packing of these regular polyhedra without segregation of the cations would necessarily be expected to produce a cubic symmetry space group. The formation of a lower symmetry space group, such as in many ternary tetragonal and hexagonal oxide structures, is brought about by segregation of the metal cations into distinct layers or columns that necessarily break spatial symmetries. Avoiding this type of segregation is explicitly required for a high entropy material with a single cation sublattice.

A further consideration that may be relevant to the over-representation of cubic symmetry HEOs is the absence of global symmetry breaking structural transitions upon cooling. To our knowledge, no such structural transitions have been observed in HEOs. We can understand why such processes may be disfavored on intuitive ground. Each ion independently may prefer an expansion, contraction, elongation, rotation, or distortion of its local oxygen environment. However, these preferences may be in direct contradiction with the preferences of the other cations in the material and therefore no net preference is exerted, and the crystal retains its high symmetry structure on the global level as a compromise. A symmetry-lowering distortion would also lower the vibrational entropy, to be discussed further below, which additionally contributes to its energetic unfavorability. 

\section{Entropy enhancements due to the local environment}

While collective symmetry reducing distortions are expected to be disfavored, the same cannot be said for non-collective distortions of the local environment. Indeed, the existence of a well-defined average crystal structure does not preclude the existence of  distortions in the local environment, which will tend to enhance the overall positional entropy of the material while lowering the enthalpy cost of an unfavorable environment. The anionic sublattice in oxide materials is particularly malleable to local distortions and defects.  

\begin{itemize}[leftmargin=*]

\item \textbf{Expansions and contractions of the local environment:} In a conventional oxide, the oxygen-cation bond distance are dictated by the ionic radius, coordination number, and oxidation state of the cation in question. In order to form a single phase HEO, each participating cation is required to make some accommodation to preserve the average global structure. Locally, however, each ion will tend to slightly isotropically distort its coordination polyhedra away from the average value to best suit its desired size, as shown schematically for a smaller or larger cation in Figure~\ref{distortions}. This scenario has been experimentally verified in the case of the prototypical rock salt HEO, using an elementally resolved x-ray absorption technique~\cite{rost2017local}. These variances do not significantly extend past the first coordination shell, with no resolvable differences being found for the second nearest neighbor bond distance. An even more extreme example comes from the orthorhombically distorted pyrochlore-type HEO Nd$_2$(Ta,Sc,Sn,Hf,Zr)$_2$O$_7$, where a neutron pair distribution function analysis revealed cation-specific distortions to the local environment so significant that the data could not be adequately modelled by the average structure~\cite{jiang2020probing}. 

\item \textbf{First order Jahn-Teller distortions:} Beyond simple expansions or contractions of the oxygen environment, cation specific electronic instabilities can also lead to \emph{anisotropic} local distortions in HEOs. One such example is the first order Jahn-Teller effect, which arises when a large electronic degeneracy can be lifted through a distortion of the local coordination environment, lowering the overall energy~\cite{goodenough1998jahn}. This effect is synonymous with octahedrally coordinated transition metals, wherein it manifests as an axial elongation or compression. It is worth emphasizing that when such effects occur in conventional oxides, they can lead to long-range symmetry-lowering structural phase transitions. In HEOs, however, they cannot occur collectively due to the dilute and random positions of the Jahn-Teller active ions. Nonetheless, the energy savings of breaking large electronic degeneracies is still substantial enough to lead to \emph{local} distortions for strongly Jahn-Teller active ions. One such example is octahedral Cu$^{2+}$ in the prototypical rock salt HEO, which has a four-fold degenerate hole in its $e_g$ orbitals, where local probes have detected a strong  axial elongation~\cite{rost2017local,zhang2019long}. Jahn-Teller distortions have also been detected for octahedral Mn$^{3+}$ and Co$^{2+}$ in the spinel HEO~\cite{JohnstoneSpinelHEO}.

\begin{figure}
  \centering
  \includegraphics[width=7.0cm]{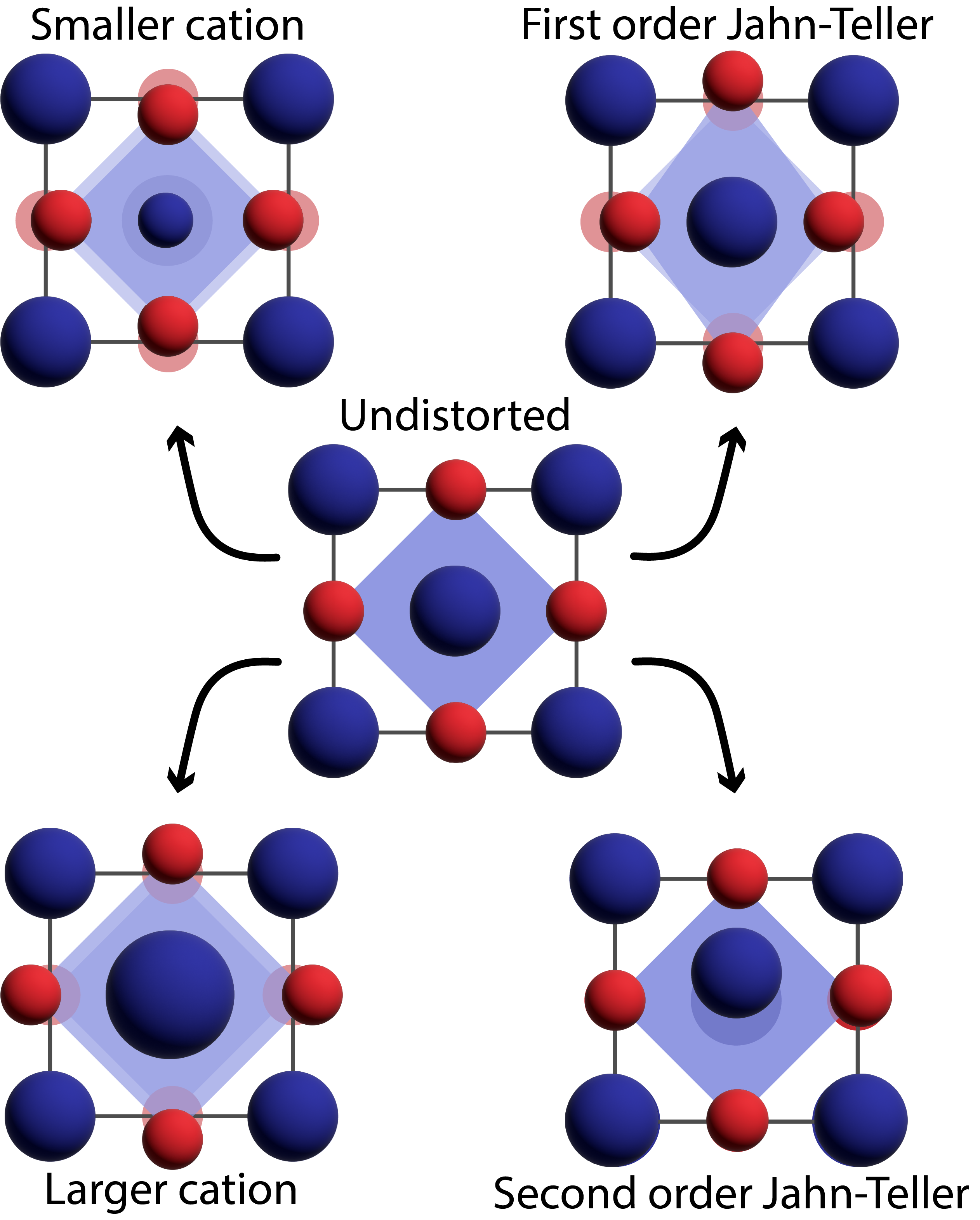}
  \caption{HEOs exhibit a well defined average undistorted structure (center) but are prone to a variety of local distortions, particularly of their oxygen sublattice. Smaller cations (upper left) and larger cations (lower left) can cause an isotropic contraction or expansion of the coordination polyhedra while Jahn-Teller active ions (upper right) can induce an anisotropic distortion to their local environment. Distortions to the cation sublattice, such as the polar off-centering of the cation due to the so-called second order Jahn-Teller distortion (lower right) are less prevalent but can also occur.}
  \label{distortions}
\end{figure}

\item \textbf{Second order Jahn-Teller distortions:} Research into local distortions in HEOs has more extensively considered the oxygen sublattice, and indeed, these are the types of distortions we would generically expect to be most prevalent. However, there is also the possibility of structural distortions to the cation lattice, such as the cation-off-centering observed in $d^0$ cations with strong second-order Jahn-Teller effects as shown schematically in Figure~\ref{distortions}. Similar to the above-mentioned first order Jahn-Teller effect, distortions of this type would not be expected to yield a macroscopic polar state in such a diluted system, but rather would occur locally on a cation specific basis. One candidate for such behavior is the high entropy pyrochlore Nd$_2$(Ti,Nb,Sn,Hf,Zr)$_2$O$_7$, where a very strong anisotropic octahedral distortion is found for Ti in a possible $d^0$ configuration~\cite{jiang2020probing}. Relaxor ferroelectrics, which lack long-range polar order, are usually highly disordered so this is an area where the concept of high entropy could lead to new applications. The first studies in this area are already showing promising results with thin films of the perovskite Ba(Ti,Sn,Zr,Hf,Nb)O$_3$ exhibiting relaxor behavior in its dielectric properties. Transmission electron microscopy has shown that the cation distortion in this material occurs collectively over a few unit cells in polar nanodomains, suggesting that not just Ti but the other $d^0$ elements are also participating in the behavior~\cite{Sharma2022}.

\item \textbf{Point defects including vacancies:} Given the profound level of configurational disorder in HEOs, more conventional forms of disorder in these materials have remained relatively unexplored. This includes especially point defects, which are local deviations from the ideal crystal structure including vacancies on the cation or oxygen sublattice, elements on interstitial sites, impurity or doping elements, and anti-site defects. Given the nature of HEOs, some of these point defects are more important to consider than others. For example, impurity  elements in small concentrations will be a small perturbation on the already large cation sublattice configurational entropy. Other types of point defects may contribute more meaningfully to the total entropy, and will generally obey the same $x \ln x$ factor. For example, a vacancy concentration of 1\% will contribute $0.01\ln(0.01)R \approx 0.05 R$, which should be compared to the ideal $S_{conf} = 1.61R$ value for the cation sublattice configurational entropy of a five-component HEO.

Vacancies are the type of point defect that has drawn the most attention in the high entropy community. In HEAs, it has been shown that the equilibrium vacancy concentration increases with configurational entropy \cite{wang2017thermodynamics}, with a five-component HEA expected to have ten times the vacancy concentration of pure metals. The same thermodynamic argument is valid for cation vacancies in HEOs. An interesting investigation on both cation and oxygen vacancies in the rock salt HEO shows that the energy of cation vacancy formation is element-specific and dependent on the local strain each cation is subjected to, and that the local environment around an oxygen vacancy determines its formation energy \cite{Chae2022}. Studies of vacancy concentrations and selectivity in oxides could help pin-point the source of their increased functionality.

\end{itemize}

We conclude this section with an extreme example of an entropy enhancement due to local distortions of the oxygen sublattice that does not neatly fit into any of the categories enumerated above. The material in question, which is the delafossite \ce{LiNiO2}, does not at first blush appear to belong in an article on high entropy materials, as both Li and Ni in this structure independently occupy their own cation sublattices. This material avoids its expected collective Jahn-Teller global symmetry lowering distortion and instead appears to exist in a glass-like structural phase. Recent work has indicated that the Jahn-Teller avoidance in this material originates from competition between different locally distorted \ce{NiO6} configurations, including three possible directions of the Jahn-Teller distortion and two size disproportionated configurations in which the entire local octahedral environment moves isotropically closer or away from the Ni cation~\cite{foyevtsova2019linio}. A near energetic degeneracy between these five configurations yields a glassy state in which all five configurations are locally present and no global symmetry reducing structural transition can proceed. The calculated ideal configurational entropy associated with this state exceeds $2R$ and therefore exceeds the high entropy threshold despite the absence of elemental configurational disorder serving as a fascinating counterpoint to the other examples discussed in this Perspective.

\section{Reductions from the ideal configurational entropy}
\label{fac-reducing-section}

\begin{figure}
  \centering
  \includegraphics[width=8.5 cm]{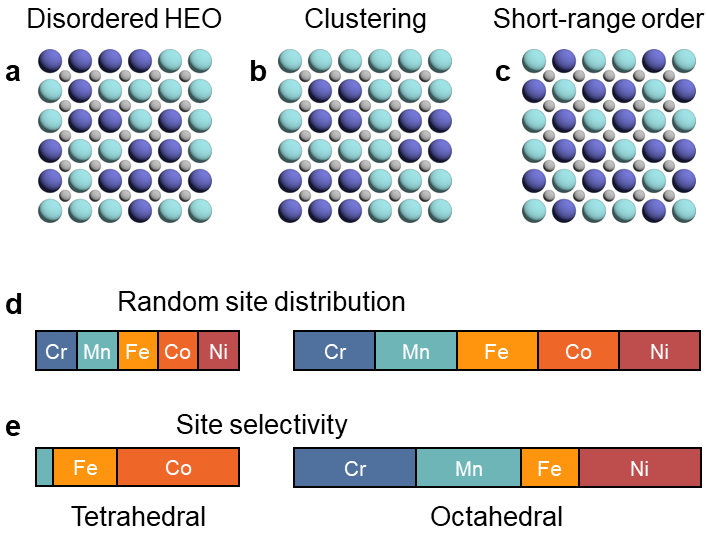}
  \caption{(a) The ideal configurational entropy in an HEO is achieved when there is a truly random distribution of cations across the shared crystallographic site. Entropy reductions can occur due to (b) clustering, wherein cations of the same type prefer to neighbor one another or (c) short-range ordering where partial ordering occurs on small length scales. (d) In materials where the cations are disordered across multiple sublattices,  the maximum entropy is achieved by a random site distribution as shown schematically for the octahedral and tetrahedral sites of a $3d$ transition metal spinel crystal structure, where there are two times as many octahedral sites as tetrahedral. (e) In the real material, crystal field effects outweigh configurational entropy leading to significant site selectivity, with the distribution redrawn from Ref.~\cite{JohnstoneSpinelHEO}.}
  \label{fgr:cluster}
\end{figure}

The configurational entropy $S_{conf}$ as calculated from Eqn.~(\ref{sid-eq}) assumes that true randomness is achieved in the distribution of the ions and that the observed distribution is but one of a very large number of fully degenerate configurations. In reality, the high entropy phase may be in close energetic competition with any number of fully or partially ordered states. As a consequence, samples that are presumed to have an ideally randomized distribution of elements may instead exhibit clustering, short-range ordering, or site selectvity which will in turn reduce the configurational entropy away from the ideal value, as shown in Figure~\ref{fgr:cluster}. We will briefly enumerate these configurational entropy reducing structural motifs and provide examples where they have been demonstrated to occur.

\begin{itemize}[leftmargin=*]

\item \textbf{Clustering and elemental inhomogeneity:}
Many of the accepted definitions for a high entropy material take an absence of clustering as an \emph{a priori} required characteristic~\cite{zhang2019review,musico2020emergent}. Indeed, the presence of clustering would generally suggest that the entropy of mixing is not sufficient to overcome the mixing enthalpy for at least one specie in the mixture, rendering that specie insoluble in the mixture and resulting in (macroscopic) phase separation. Such a scenario can indeed be observed in some reported HEO 
materials, such as the fluorite (Ce,Zr,Hf,Sn,Ti)O$_2$, where elemental mapping shows segregation of Ce into micron scale domains~\cite{chen2018five}. However, in coming to grips with what such clustering means in terms of the materials acceptance under the HEO umbrella, it is worth taking into consideration both the length scale and the origin of the inhomogeneity. It may be the case that the homogeneous entropic state \emph{is} thermodynamically preferred at the synthesis temperature, but slow reaction kinetics prevent a complete transformation. In such cases, synthesis method will play a key role in determining the observed homogeneity, with techniques that yield a higher initial degree of randomness giving superior results. In the case of the spinel HEO, (Cr,Mn,Fe,Co,Ni)$_3$O$_4$, for instance, samples prepared via spray pyrolysis exhibit far higher levels of elemental homogeneity at the micron scale~\cite{sarkar2022comprehensive} than samples prepared via solid-state synthesis~\cite{JohnstoneSpinelHEO}, for which the initial mixing is poorer. Another case where synthesis conditions are the culprit is for the prototypical rock salt HEO (Mg,Co,Ni,Cu,Zn)O, wherein synthesis at either too high or too low temperature results in clustering and segregation of CuO or Cu$_2$O~\cite{hong2019microstructural,dupuy2019entropic}. 

\item \textbf{Short range ordering:} The intrinsic mixture of ionic sizes, valences, and even magnetic states of HEOs will all tend to promote short-range order between the elements, even if only on a hyper-local length scale. Due to the strong and long-ranged electrostatic forces, a material which globally looks completely disordered may have strong local ordering of charge. Quasi-ordering on very small length scales is unlikely to manifest itself in average probes, such as x-ray diffraction, nor would it be easily detected in real space microscopies unless the domain size grows to a nanometric scale. Yet, the reductions to entropy arising from favorable pairwise correlations between two or more constituents in a HEO are real and will lead to significant reductions to the entropy from its ideal value. Even in compositionally `simple' oxides, detecting short range ordering at the nanoscale is a  challenging endeavour~\cite{keen2015crystallography,OQuinn2020}. Within the HEO literature, this topic has not yet been extensively tackled, although it has been identified by many as an area for expanded research effort~\cite{musico2020emergent,brahlek2022name}. One example comes from the compositionally simple but entropy stabilized (Ti,Zr)O$_2$ in the $\alpha$-PbO$_2$ structure. This material has a near random arrangement of Ti and Zr when quenched from high temperatures but exhibits short-range Ti/Zr ordering in samples that are more slowly cooled due to incipient long-range ordering hindered by very long equilibration times.

\item \textbf{Site selectivity:} A unique circumstance that can arise when the structure contains more than one crystallographically unique cation site is site selectivity. In certain materials, such as the high entropy perovskite and pyrochlores~\cite{sarkar2018rare,Li2019}, almost perfect site selectivity is expected due to the large differences in coordination number and the resulting volume of the cation site, such that there are large energetic penalties to putting, for example a rare earth ion on the pyrochlore lattice $B$-site. In these cases, the largest configurational entropy is not achieved, which would occur when there is a perfectly random distribution of all cations across both cation sites. In other structure types, such as the spinel structure, the size of the two cation sites is more comparable and here a higher degree of mixing can be expected. Indeed, even in conventional spinels, such as \ce{MgAl2O4} entropy-driven site mixing is observed. In this material, size arguments alone lead to the enthalpically selected ground state with Mg occupying the tetrahedral site and Al preferring the octahedral site. However, quenching from different synthesis temperatures can be used to tune the level of inversion due to the enhanced configurational entropy~\cite{schmocker1976inversion}. This picture gets even more complicated once there are more ions involved each with a specific degree of site preference, and especially when $d$ electrons with crystal field effects come into play. For instance, a cation distribution far from the entropic ideal is indeed observed in the case of the spinel HEO (Cr,Mn,Fe,Co,Ni)$_3$O$_4$, which is driven by crystal field effects, as shown by the cation distribution in Figure~\ref{fgr:cluster}(d,e)~\cite{sarkar2022comprehensive,JohnstoneSpinelHEO}.
\end{itemize}

\section{Non-configurational contributions to entropy}

\label{Contributions-to-S}

Although the configurational entropy discussed previously is typically considered as the main contributor to phase stabilization and property enhancement in high entropy materials, non-configurational contributions to the entropy do in some cases dominate the thermodynamics. In the following sections, the notion of entropy will be extended to vibrational ($S_{vib}$), electronic ($S_{el}$), magnetic ($S_{mag}$), orbital ($S_{orb}$), and charge ($S_{charge}$) degrees of freedom. 
The different sources of entropy discussed in this section have variable magnitudes in different temperature ranges. Boltzmann's entropy formula in statistical mechanics, $S = k_B \ln \Omega$, embodies the notion of entropy in the number of microstates the system is exploring over a temporal or spatial average. Thus, while we can express the total entropy $S_{total}$ as
\begin{equation}
    S_{total} = S_{conf} + S_{vib} + S_{el} + S_{mag} + S_{orb} + S_{charge}...
\end{equation}
it is important to remember that the relevance and magnitude of these different terms will depend strongly on the temperature scale under consideration. Furthermore, the microstates accessible for a given term in the entropy may be limited by another term that was resolved at a higher temperature, creating an effective hierarchy. When a high-entropy material is cooled down and the disordered configuration is frozen in the lattice, the system is no longer dynamically exploring all possible microstates, yet the entropy is not lost since all possible microstates are still represented in a spatial average \cite{Takada2015}. This implies that a quenched high-entropy oxide is out of thermodynamic equilibrium and should exhibit residual configurational entropy all the way down to 0 K, in analogy to the residual entropy often considered in reference to the frozen positional disorder of glasses~\cite{simon1937third} or the frozen degenerate magnetic disorder in spin ice \cite{Bramwell2001}. 

\subsection{Vibrational entropy}
 
Vibrational entropy arises from small, almost perfectly harmonic excitations of the atoms about their equilibrium positions and is known, in some cases, to stabilize high temperature phases~\cite{fultz2010vibrational}. In order to calculate the vibrational entropy from Boltzmann's formula, consider the number of ways for $m$ phonons to occupy $M$ oscillator states which is the number of microstates $\Omega=(m+M)!/m!M!$, and the occupancy of a state is $n=m/M$. The vibrational entropy is obtained as $S_{vib}=(1+n) k_B \ln(1+n)-n k_B \ln n$. Since phonons are bosonic, their temperature dependence obeys Bose-Einstein statistics, $n(\epsilon)=[\exp[\epsilon/k_BT]-1]^{-1}$. Using the vibrational density of states $g_{vib}(\epsilon)$, the vibrational entropy within the harmonic approximation can be written as an integral,
\begin{equation}
    S_{vib}=3k_B\int d\epsilon g_{vib}(\epsilon)[(1+n(\epsilon))\ln(1+n(\epsilon))-n(\epsilon)\ln n(\epsilon)]
\end{equation}
Importantly, the only material parameter needed for the calculation of $S_{vib}$ is the phonon density of states. While high entropy materials do have a well-defined average crystal structure, which in principle would imply a well-defined phonon density of states, the realities outlined in previous sections make the picture more complicated. Starting from a simplistic balls and springs model used in the classical treatment of phonons, it is immediately clear that while it is possible to define an average spring stiffness or an average mass on the cation site, there will be large dispersions across both values across the solid. Consequently, significant broadening in phonon modes and highly shortened lifetimes due to scattering can both be expected in high entropy materials. This phonon broadening has been experimentally verified in a handful of oxides with Raman spectroscopy~\cite{sharma2018single,dkabrowa2018synthesis}. As expected, the thermal conductivities of HEOs are substantially reduced from their non-configurationally disordered analogs~\cite{braun2018charge,sharma2018single,chen2018five,lim2019influence}, in some cases even approaching the amorphous limit. 

Setting aside the precise form of the phonon density of states in HEOs, we turn to the question of whether vibrational entropy plays a role in their formation. To our knowledge, there are no studies that address this question, but some parallels can be drawn from studies on alloys and carbides. In HEAs, the vibrational contribution to the free energy can exceed the configurational term at room temperature and overshadow it at typical synthesis temperatures~\cite{Ma2015entropybeyondconfigurational}. However, the expected change in the vibrational entropy $\Delta S_{vib}$ associated with the transformation from elemental metals into a single phase alloy is usually smaller than the configurational contribution as the competing phases typically have similar bonding and vibrational entropies \cite{HighEntropyAlloysBook}. A typical magnitude for $\Delta S_{vib}$ can be a few kJ/mol-K for HEAs as compared to the change in configurational entropy $\Delta S_{conf}$ for a five component system which is an order of magnitude larger. Notably, the sign of $\Delta S_{vib}$ can be either positive or negative, destabilizing or stabilizing the high entropy compound in question.

In high entropy carbides, vibrational contributions become important when precursors or decomposition products have different nearest-neighbor environments~\cite{esters2021settling}, which we expect to carry over to HEOs. In both carbides and alloys, phonon frequencies decrease with increasing volume or higher masses leading to higher vibrational entropy. It is also universal that mass disorder destabilizes the high entropy compound whereas force constant disorder stabilizes it~\cite{esters2021settling, Kormann2017phononbroadening}. The vibrational entropy might not be the deciding factor in whether or not a high entropy phase will form, but it can change the transition temperature by hundreds of Kelvin (thus affecting the kinetics), it can change the size of the miscibility gap where synthesis is possible, and it could be a factor in choosing the observed polymorph. There is a clear imperative to perform similar studies on HEOs through the lens of vibrational entropy to understand its role in stabilizing the high entropy phase.

\subsection{Electronic entropy}
In insulating oxides, such as all the HEO materials discussed here thus far, all electrons are highly localized and there is therefore no contribution to entropy arising from the occupation of thermally excited electronic energy levels. However, in terms of their electronic properties, oxides are not a monolith: many oxides based on $4d$ and $5d$ transition metals are semiconducting or metallic. To our knowledge, there are no reports of metallic HEOs where the conduction band is associated with the configurationally disordered sublattice. However, they are a small number of reports of metallic HEOs where the conduction occurs in bands associated with a uniform cation ordered sublattice, which is only weakly perturbed by the mixture of cations on the secondary sublattice. A particularly interesting example is (La,Pr,Nd,Sm,Eu
)NiO$_3$, grown as epitaxial films, which exhibits a first order metal-to-insulator transition~\cite{patel2020epitaxial}.

To determine the electronic contribution to the entropy we can directly modify our previous equation for configurational entropy, Eq.~\eqref{sid-eq} where the fraction $x_i$ can be replaced by a "concentration" of electrons with a specific energy $\epsilon$. Since electrons are fermions, their temperature dependence is described by the Fermi-Dirac distribution $f(\epsilon)=[\exp[(\epsilon-\epsilon_f)/k_BT]+1]^{-1}$, where with increasing temperature their distribution gets smeared out around the Fermi energy. With continuous energy levels and an electronic density of states $g_{el}(\epsilon)$, the electronic contribution to the entropy can be written as
\begin{equation}
    S_{el}=-k_B\int d\epsilon g_{el}(\epsilon)[f(\epsilon)\ln f(\epsilon)+(1-f(\epsilon))\ln(1-f(\epsilon))]
\end{equation}
The only material parameter is the electronic density of states, with the most importance given to the density around the Fermi level. For insulating oxides, $S_{el}$ is negligible as there are no states near the Fermi level. However, if a metallic HEO were discovered we can anticipate, through comparisons with the HEAs, a typical magnitude for the electronic entropy of around 1 J/K/mol at 1000 K\cite{HighEntropyAlloysBook}. The electronic contribution to the entropy is therefore orders of magnitude smaller than the vibrational contribution even in cases where both the competing and  resulting phases are good electronic conductors with large densities of states around the Fermi level. It is therefore highly unlikely that the electronic entropy would ever come in to play for determining the entropy stabilized phase. However, in the event of phase competition between phases where some are insulating and others are conducting, the electronic entropy would weakly favor the conducting phase.

\subsection{Orbital, spin, and charge entropy}

Every electron has two degrees of freedom, the orbital occupation and the spin state (up or down). The multiplicity of unpaired electrons with particular spins in specific orbitals can be calculated in a similar fashion to the configurational entropy in Eqn.~\ref{sid-eq}. For example, a single electron in a fully degenerate $t_{2g}$ orbital would have an entropy of $S_{mag}=-\ln(6)R=1.79R$. This entropy contribution is therefore large in the paramagnetic high temperature state typically found at the synthesis temperature, but greatly reduces or completely vanishes in a magnetic ordering transition or in a Jahn-Teller distortion further lifting the degeneracy. Additionally, the competing phases will often have similar magnetic and orbital characteristics leading to a smaller entropy of reaction, similar to the vibrational entropy. Coexistence of high-spin and low-spin states of the same element can create an additional contribution to the entropy, and the same is true for multiple coordination environments or multiple valence states. For example, in the high entropy spinel (Cr,Mn,Fe,Co,Ni)$_3$O$_4$, Mn is observed to partially occupy both the tetrahedral and octahedral sublattices in a mixture of 2+, 3+, and 4+ oxidation states, each with its own associated magnetic entropy. Each of these oxidation states will contribute to the configurational entropy separately, again with an $x\ln x$ factor.

\section{What is the role of kinetics in a high entropy material?}
\label{sec-kinetics}

\begin{figure*}
  \centering
  \includegraphics[width=15 cm]{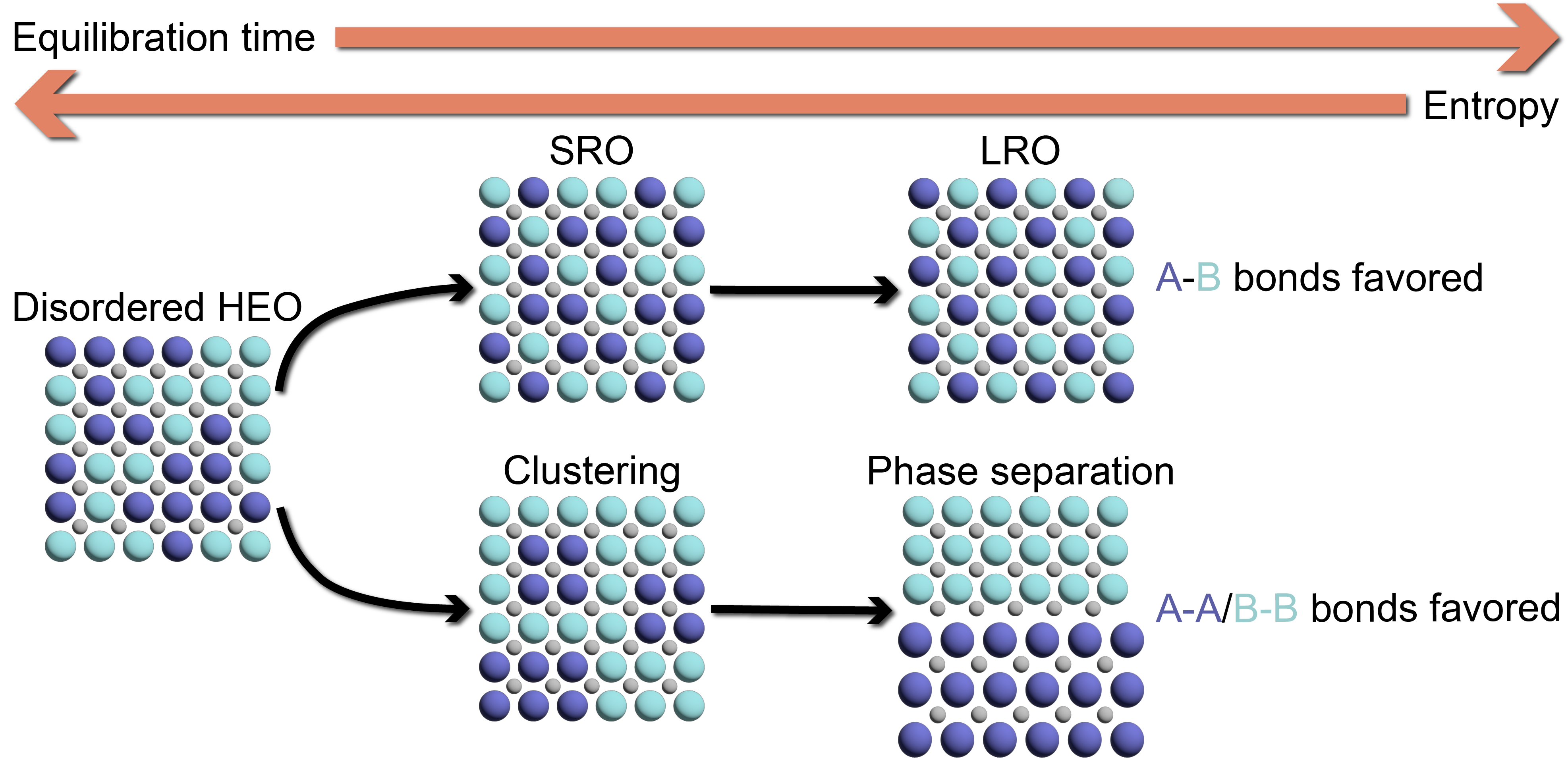}
  \caption{The thermal history a sample has been exposed to and the energetics of the particular system decides which final state the sample exhibits. A disordered HEO with a high enthalpy and high entropy can be retained by quenching, while longer equilibration times allows for the formation of long range order (LRO) or complete phase separation. Intermediate equilibration time will cause clustering or short range ordering (SRO).}
  \label{fgr:kinetics}
\end{figure*}

High-entropy oxides are typically synthesized at high temperatures with long reaction times, and as such it is commonly assumed that thermodynamic equilibrium is reached at the synthesis temperature, and that subsequent quenching from the synthesis temperature retains the completely disordered phase at room temperature. However, there are indications that the influence of kinetics may be profound in these materials and, as a result, the structure and functional properties of the final reaction product may depend sensitively on synthesis method and cooling rates.

Historically, the interplay between kinetic and thermodynamic control of a reaction product has been a more commonly used and well-understood tuning parameter in the field of organic chemistry than in inorganic or solid state chemistry. If there are at least two possible reaction products, the key to choosing one reaction product is realizing that the one with lower activation energy will form first (kinetic control) while the one with the most favorable Gibbs free energy will form if given sufficient time and energy to do so (thermodynamic control) \cite{Marsden2020}. By optimizing reaction time and temperature, the reaction can be engineered to yield the desired product. Given their presumed metastable nature and large number of competing phases, there are clear implications for the formation of high entropy materials.  

Kinetics do indeed play a role during the formation of even conventional oxide phases. Upon heating, the reactants may go through several out-of-equilibrium steps of metastable phases before arriving at the target compound, as measured by in-situ diffraction during solid-state synthesis of layered sodium cobaltite magnetite oxides \cite{Bianchini2020}, or hydrothermal synthesis of sodium niobate \cite{Skjarvo2017}. The initial mixing of the reactants before synthesis could determine which phase is able to form first, which might be a reason why solution-based methods such as nebulized spray pyrolysis and solution combustion are prominent in high entropy materials synthesis, since they usually promote more intimate initial mixing of the constituents. Equally popular solid state methods on the other hand, rely on the diffusion of individual atoms over length scales upwards of hundreds of nanometers depending on the particle size of the initial reagent. In such a case, an insufficiently homogenized preparation might be expected to lead to the kinetically driven formation of impurity phases involving a subset of the reagents, and these intermediate phases could become trapped. Along similar lines, it is possible in such a case that the reaction proceeds to a single phase but that elemental inhomogeneity remains due to diminishing returns of long-range diffusion once the sample is close to, but not in, thermodynamic equilibrium. Alternate synthesis methods for HEOs, such as pulsed laser deposition (PLD)~\cite{Kotsonis2018epitaxialkinetic} and carbothermal shock synthesis~\cite{yao2018carbothermal}, provide extremely rapid quenching to the point where kinetic control is achieved, or mechanochemical synthesis which avoids heat entirely allowing the inclusion of redox sensitive cations and multi-anionic oxyfluorides \cite{Lin2020mechanochemical}. These methods allow the synthesis of out-of-equilibrium phases that are inaccessible with solid-state methods and are therefore excellent routes for the discovery of new HEO phases.

Figure \ref{fgr:kinetics} summarizes what the result of cooling an entropy stabilized binary compound at different rates could be. If cooled sufficiently quickly, the material stays in its high entropy, high enthalpy, completely configurationally disordered HEO state. This option has no activation barrier, and can be seen as the kinetically controlled reaction product. With longer equilibration times at elevated temperatures, we can end up in different states depending on the atom-atom interactions. If the atom-atom interactions of the mixture favor A-B bonds over A-A/B-B bonds, the material is expected to form a long range ordered (LRO) compound given sufficient equilibration time, conversely, if A-A/B-B bonds are favored, complete phase separation would be expected. One of these two options might be the most thermodynamically stable, or the thermodynamically controlled reaction product, but depending on the height of the energy barrier and the cooling rate, this state might not be accessible.
Between these extrema, intermediate states where the structure reorganizes to some degree are imaginable, where either clustering or short-range ordering as discussed in section IV could be observed. 

There are some general trends in the height of kinetic energy barriers, generally the long-range diffusion required for phase separation or long-range order will have the highest energy barrier. The two options have the same barrier energy, but LRO is more kinetically accessible than phase separation due to the shorter required diffusion lengths. SRO and clustering require even shorter diffusion lengths and hence are more kinetically accessible. Oxygen vacancies typically have higher diffusion rates than cation vacancies, and diffusion in alloys are generally faster than in oxides. Diffusionless transitions will be faster than those requiring diffusion, and within those types of transitions the reconstructive transitions are typically slower than displacive transitions. Finally, magnetic and electronic transitions have the lowest energy barriers and might not be frozen in even below room temperature.

While the influence of kinetics in HEOs have largely been overlooked until now, there are several striking examples showing that the thermal history of a sample can affect the final structure and functional properties. Quenching can stop the diffusion of ions, while a slow cooling rate or long annealing time can allow for the transition to a thermodynamically more stable configuration. Taking (Ti,Zr)O$_2$ as an example, by quenching the sample from a high temperature, a disordered distribution of Ti and Zr ions are retained at room temperature; while using slow cooling rates, the Ti and Zr cation reorganize in slabs along the b-axis, forming a superstructure \cite{Christoffersen1992}; and finally, using extensive annealing times (four months) just below the formation temperature, a partial dissolution of the competing phases is observed \cite{Mchale1986}. Another example of the intimate relation between atomic configurations and thermal history is found in the prototype rock-salt system, where different cooling rates and annealing times and temperatures influence the displacive Jahn-Teller distortion of Cu~\cite{berardan2017controlled}. Kinetic effects are also an important consideration when it comes to the functional properties of high entropy material, as exemplified by the changes in dielectric properties observed in the rock-salt HEO~\cite{Berardan2017}. Beyond oxides, quenching and annealing time have also been used to tailor the phase and multiphase composition in high entropy alloys \cite{He2016}.

There is also a strong interplay of reaction kinetics with reversibility, which is a particularly important consideration given that reversibility has been held up as a key indicator of entropy stabilization, as discussed previously. However, reversibility depends on long-range diffusion of cations, and might be kinetically hindered if the theorized transition is at a point where the cation diffusion rate is limited. Even in cases where only low levels of disorder are present, the reversible phase transition of an entropy stabilized conventional oxide can be kinetically hindered or even completely trapped, as has been shown for PbNb$_2$O$_6$ \cite{Roth1957,Sahini2014}. Given the observed kinetic hindrance in systems with low levels of disorder the macroscopic configurational disorder in high entropy materials can certainly be expected to have a profound kinetic effect on the likelihood of observing the reversibility of their formation. While reversibility when present in a specific high entropy material is indeed evidence of entropy stabilization, it should not be used as an absolute requirement for something to be called entropy stabilized. Reversibility is inextricably entangled with kinetics, while entropy stabilization can be thought of as an equilibrium thermodynamic concept.

\section{Outlook}
This perspective has highlighted many aspects of HEOs that require a deeper understanding of the role of entropy - both for their stability and for their functional properties. Only a tiny fraction of the huge space of possible chemistries has been explored so far. The gold standard of conventional solid state research is when experimental reality can be well described by a theoretical model, ideally computed \textit{ab initio}. Despite the existence of a well-defined crystal structure, high entropy materials push this paradigm to a new extreme. Within density functional theory (DFT) calculations, HEOs can be approached using special quasirandom structure (SQS) supercells as specific disorder realizations, from which formation enthalpies can be calculated and information about the local charge state be obtained \cite{rak2016charge}. However, rapid computational screening of a large number of element combinations is not possible with the SQS approach due to the large computational overhead. A computationally efficient strategy introduced recently suggests the use of nearest neighbor pairwise approximations with interactions computed in small unit cells comprising a subset of the elements at a time \cite{pitike2020predicting,jiang2020probing}. This method makes predictions for the relative stability of hypothetical compounds based on entropy and enthalpy descriptors. 

Direct computation of the multiple contributions to the entropy would be desirable, but comes with challenges. The electronic and vibrational entropies can be assessed fairly quickly from standard routines using the electronic and phononic band structures \cite{hossain2021entropy}. These techniques have already found extensive use in the field of high entropy alloys \cite{HighEntropyAlloysBook}. Obtaining reliable estimates for the configurational entropy is nontrivial. As mentioned before, Eq.~(\ref{sid-eq}) applies only to a truly random solid solution and thus provides an upper bound of $S_{conf}$ at high temperature. Addressing local effects discussed in Section \ref{fac-reducing-section} is not possible with DFT calculations at zero temperature alone. Finite temperature lattice Monte Carlo simulations can be used to explore this configuration space, but require an extrapolation of the \textit{ab initio} interaction energies. One possible approach is the cluster expansion, which expands the energy systematically into multibody interactions and fits the expansion parameters to DFT data \cite{chang2019clease}. An alternative, not yet fully explored possibility is to employ machine learned interaction potentials (MLIPs) \cite{friederich2021machine} for efficient computation of the energy differences of Monte Carlo moves. Clustering and other short range ordering are emergent in such simulations, and the entropy can also be calculated directly. When coupled with experimental techniques at the smallest length scales such as atom probe tomography, EXAFS, and PDF analysis, such an approach might be the most promising route to directly quantifying the true entropy in this class of materials. 

The next computational crux in the high entropy field is the prediction of functional properties. Accurate predictions of this kind is difficult enough for conventional ordered compounds. Emergent properties - where a mixture exhibits functionality which is greater than the sum of its constituents - are at the core of what makes these materials attractive. This implies that the functional property prediction has to deal with the full chemical complexity of the system, potentially also with complicating factors such as short range ordering and non-equimolar stoichiometry. Consequently, there are few instances of calculated or predicted properties, some examples are the calculation of electronic properties \cite{Zywczak2020} and adsorption energies \cite{Feng2020}. There is also some promise in predicting mechanical properties from machine learning \cite{Tang2021}. MLIPs for high entropy alloys have been used to screen the temperature and composition dependence of elastic constants identifying alloys displaying invariable elasticity~\cite{Gubaev2021mlip}. Given thant the applications typically mentioned for HEOs are thermoelectrics and ion conductors, enabling high-throughput calculations of electronic, thermal, and ionic transport properties is perhaps the most rewarding route.

The computational front of high entropy materials is generally lagging behind the experimental front. Chemical intuition is sufficient to guess which constituent oxides will form a single phase, and mixing oxides together in the lab by trial and error is in our experience still faster than running predictions \textit{in silico}. Considering the enormous potential parameter space of high entropy materials, a huge influx of new compounds is to be expected. The goal of the computational work in this field might be to determine what should be synthesized, not just what can be synthesized.

While this Perspective has emphasized understanding the role of entropy in HEOs, it is important to consider how HEOs fit in the broader landscape of high entropy materials. The field of HEOs is younger and, for now, smaller than that of HEAs: in 2022 the Web of Science Core Collection had 2845 entries for HEAs and 748 for HEOs. While many concepts from HEAs can be directly applied to the oxides there are some key differences separating the oxides from the alloys. While local distortions in high entropy alloys can be largely understood as originating from differing sizes of the atoms, the problem is further complicated in oxides due to the mediating effect of the oxygen sublattice and intricacies of charge balancing between the anionic and cationic sublattice.
We anticipate that the phenomenological description of HEOs will ultimately be quite distinct from that of alloys.

Beyond oxides and alloys, there are also burgeoning fields based on other materials families, such as high entropy borides~\cite{gild2016high}, carbides~\cite{yan2018hf0,castle2018processing,sarker2018high}, nitrides~\cite{jin2018mechanochemical}, chalcogenides~\cite{zhang2018data,jiang2021high}, and fluorides~\cite{wang2020high}, as well as different modalities such as two-dimensional van der Waals bonded materials~\cite{ying2021high} and nanoparticles~\cite{wang2022general}. Materials with disordered anion sublattices, both as the primary or secondary source of configurational entropy, are also an emerging area of interest. Each of these materials classes presents their own distinct opportunities and challenges. These interconnected fields should continue to draw inspiration from one another towards the unifying goal of understanding, what is the true role of entropy?

\begin{acknowledgments}

The authors thank the members of the Quantum Matter Institute's ``Atomistic approach to emergent properties of disordered materials'' Grand Challenge for insightful conversations on high entropy oxides. This work was supported by the Natural Sciences and Engineering Research Council of Canada and the CIFAR Azrieli Global Scholars program. This research was undertaken thanks in part to funding from the Canada First Research Excellence Fund, Quantum Materials and Future Technologies Program. 

\end{acknowledgments}

\bibliography{bibliography}

\end{document}